\documentclass[twoside,epsf]{article}
\usepackage{graphicx}
\newcommand{\lascia}[1]{}

\newcount\Mac  \Mac=0  
\newcommand{\ifMac}[2]{\ifnum\Mac=1 #1 \else #2 \fi}
\newcommand{\NO}{\hbox{---}}

\def\Red  {}

\def\Black{}
\def\Blue {}

\newcommand{\GeV}{\,{\rm GeV}}
\newcommand{\TeV}{\,{\rm TeV}}

\newcommand{\One}{\hbox{1\kern-.24em I}}
\newcommand{\NP}{Nucl. Phys.}
\newcommand{\PRL}{Phys. Rev. Lett.}
\newcommand{\PL}{Phys. Lett.}
\newcommand{\PR}{Phys. Rev.}

\newcommand{\eq}[1]{~(\ref{eq:#1})}

\def\Ord{{\cal O}}  \def\SU{{\rm SU}}
 
\def\circa#1{\,\raise.3ex\hbox{$#1$\kern-.75em\lower1ex\hbox{$\sim$}}\,}
\makeatletter
\def\art{\@ifnextchar[{\eart}{\oart}}
\def\eart[#1]#2#3#4#5#6{{\rm #2}, {\em #3 \bf #4} {\rm (#6) #5} ({\em #1})}
\def\hepart[#1]#2{{\rm #2, \em#1}}
\newcommand{\oart}[5]{{\rm #1}, {\em #2 \bf #3} {\rm (#5) #4}}

\newcounter{alphaequation}[equation]
\def\thealphaequation{\theequation\hbox to
0.6em{\hfil\alph{alphaequation}\hfil}}
\def\eqnsystem#1{
\def\@eqnnum{{\rm (\thealphaequation)}}
\def\@@eqncr{\let\@tempa\relax \ifcase\@eqcnt \def\@tempa{& & &} \or
  \def\@tempa{& &}\or \def\@tempa{&}\fi\@tempa
  \if@eqnsw\@eqnnum\refstepcounter{alphaequation}\fi
\global\@eqnswtrue\global\@eqcnt=0\cr}
\refstepcounter{equation} \let\@currentlabel\theequation \def\@tempb{#1}
\ifx\@tempb\empty\else\label{#1}\fi
\refstepcounter{alphaequation}
\let\@currentlabel\thealphaequation
\global\@eqnswtrue\global\@eqcnt=0 \tabskip\@centering\let\\=\@eqncr
$$\halign to \displaywidth\bgroup \@eqnsel\hskip\@centering
$\displaystyle\tabskip\z@{##}$&\global\@eqcnt\@ne
\hskip2\arraycolsep\hfil${##}$\hfil& \global\@eqcnt\tw@\hskip2\arraycolsep
$\displaystyle\tabskip\z@{##}$\hfil
\tabskip\@centering&\llap{##}\tabskip\z@\cr}
\def\endeqnsystem{\@@eqncr\egroup$$\global\@ignoretrue} \makeatother

\newcount\n

\begin{document}\twocolumn[
\centerline{hep-ph/0007265 \hfill IFUP--TH/2000--22 \hfill SNS-PH/00--12}
\vspace{5mm}

\Black
\vspace{0.5cm}
\centerline{\LARGE\bf\Red The `LEP paradox'}\Black

\medskip\bigskip\Black
\centerline{\large\bf Riccardo Barbieri}\vspace{0.2cm}
\centerline{\em Scuola Normale Superiore, Piazza dei Cavalieri 7, I-56126 Pisa, Italy and INFN}
\vspace{3mm}
\centerline{\large\bf Alessandro Strumia}\vspace{0.2cm}
\centerline{\em Dipartimento di Fisica, Universit\`a di Pisa and INFN, Pisa, Italia}
\vspace{1cm}
\Blue\centerline{\large\bf Abstract}
\begin{quote}\large\indent
Is there a Higgs? Where is it? Is supersymmetry there?
Where is it?
By discussing these questions, we call attention to the `LEP paradox', which is how
we see the naturalness problem of the Fermi scale after a decade of electroweak
precision measurements, mostly done at LEP.
\end{quote}\Black
\vspace{0.5cm}]

\noindent
Is it wise to spend time in reviewing a subject, which can be summarized in one sentence:
neither the Higgs nor supersymmetry have been found so far?
Admittedly the question makes sense.
For sure these topics are crucial to the central problem of particle physics:
the ElectroWeak symmetry breaking.
Our main motivation here is, however, more specific.
We want to bring the focus on what we like to call the `LEP paradox'.
By this we mean the way several years of (mostly) LEP results~\cite{LEP} make us see the old and well
known naturalness problem of the Fermi scale.

The questions we address, in logical order, are:
\begin{enumerate}
\item Is there a Higgs?
\item If yes, where is it?
\item Is there supersymmetry?
\item If yes, where is it?
\end{enumerate}
All of these questions, as well as their answers, have to do with the `LEP paradox'
that was just mentioned.

\section{Is there a Higgs?}
Any decent theory of the EW interactions must contain the Goldstone bosons,
two charged and one neutral, that provide the longitudinal degrees of freedom for the $W$ and $Z$ bosons.
On top of them, the Standard Model has a neutral Higgs boson.
Without the Higgs and without specifying what replaces it, one deals with a gauge Lagrangian
with $\SU(2)_L\otimes{\rm U}(1)$ non-linearly realized in the Goldstone boson sector.
This is in formal analogy with the chiral $\SU(3)_L\otimes\SU(3)_R$ of strong interactions in the
pseudoscalar octet sector.

The predictive power of such a non-linear Lagrangian is reduced with respect to the SM.
In practice, at present, the comparison can be made by considering 2 ``$(g-2)$-like'' quantities,
$\epsilon_1$ and $\epsilon_3$~\cite{ABC}, which include the EW radiative correction effects more sensitive to the Higgs
sector.
The experimentally determined $\epsilon_1$ and $\epsilon_3$~\cite{LEP}, mostly by $\Gamma_Z$, $M_W/M_Z$ and $\sin^2\theta_{\rm W}$,
are shown in fig.~1 and compared with the SM prediction.
All the radiative corrections not included in $\epsilon_1$ and $\epsilon_3$, less sensitive
to the Higgs mass, are fixed at their SM values.
The agreement with the SM for a Higgs mass below the triviality bound of about $600\GeV$ is remarkable and constitutes indirect
evidence
for the existence of the Higgs boson.

With a non linear Lagrangian, neither $\epsilon_1$ and $\epsilon_3$ can be computed.
Some believe, however, that suitable models of EW symmetry breaking may exist where both $\epsilon_1$ and $\epsilon_3$
deviate from the SM values for $m_h=(100\div 200)\GeV$ by less than $(1\div 2)10^{-3}$, having therefore a chance of
also reproducing the data without an explicit Higgs boson in the spectrum~\cite{x}.
In the case where a reliable estimate can be made, technicolour models with QCD-like dynamics, this is known
not to happen~\cite{noTechni}.

\begin{figure}[t]
\begin{center}
\includegraphics{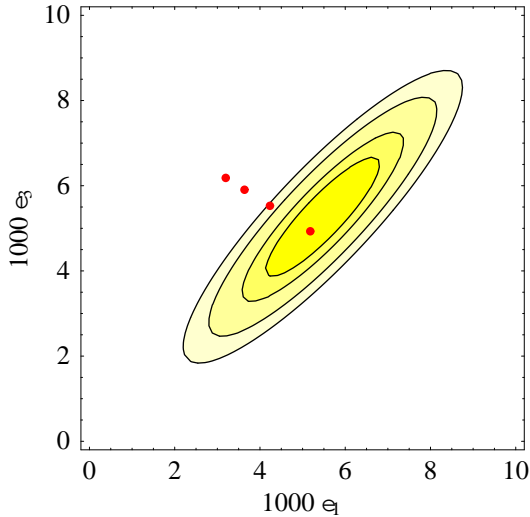}
\caption[SP]{\em Level curves at $\{68\%,90\%,99\%,99.9\%\}{\rm CL}$ of $\epsilon_1$ and $\epsilon_3$
compared with the SM prediction for $m_h=100,300,600,1000\GeV$, from right to left.}
\end{center}\end{figure}

\section{Where is the Higgs?}
If one accepts the existence of a Higgs boson, the SM Lagrangian ${\cal L}_{\rm SM}$ becomes an unavoidable effective low energy
approximation of any sensible theory.
A deviation from it could occur for the need of describing new degrees of freedom with mass comparable or lower than the
Fermi scale.
Barring this possibility, the predictions of the SM --- hence the indirect determination of $m_h$ from the EW Precision Tests ---
could only be altered by the presence of operators ${\cal O}_i^{(4+p)}$ of dimension $4+p\ge 5$ weighted by inverse 
powers of a cut-off scale, $\Lambda$, associated with some kind of new physics
$${\cal L}_{\rm eff}(E<\Lambda) = {\cal L}_{\rm SM}+ \sum_{i,p} \frac{c_i}{\Lambda^p}{\cal O}_i^{(4+p)}.$$
It is unavoidable that the ${\cal O}_i$ respect gauge invariance.
For the purposes of the following discussion, it is conservative that we restrict them to be
flavour universal and $B$, $L$, CP conserving.

How does this modified Lagrangian compare with the most recent EWPT~\cite{NRO-SM-FIT}?
Table~1 gives a list of the (independent) operators that affect the EWPT, together with the lower limits that the
same EWPT set on the corresponding $\Lambda$ parameters.
We take one operator at a time with the dimensionless coefficients $c_i=+1$ or $c_i=-1$ and different values
of the Higgs mass.
The blanks in the columns with $m_h=300$ or $800\GeV$ are there because no fit is possible, at $95\%$ C.L.,
for whatever value of $\Lambda$.
A fit is possible, however, for $m_h=(300\div 500)\GeV$ with {\em suitable\/} operators and with $\Lambda$
{\em in a defined range\/}~\cite{HK},
as shown in fig.~2.

For this reason one is cautious about saying that the Higgs is between 100 and 200 GeV, as obtained in a pure SM fit
with $\Lambda=\infty$.
To fake a light Higgs, however, a coincidence is needed.
From table~1, a more likely conclusion seems that $\Lambda$ is indeed bigger than about $5\TeV$ and the Higgs is light.

In spite of this, an interesting question is the following~\cite{HK}: suppose that a Higgs heavier than $250\GeV$ appears
light in the EWPT because of a suitable operator in the above list.
Would any direct effect of such operator be visibile in high energy collisions?
Not possible, we think, at the {\sc Tevatron} and unlikely at the LHC, where it would be definitively easier to
discover the Higgs itself.

\begin{table*}[t]
$$
\begin{array}{rll|cc|cc|cc}
\multicolumn{3}{c|}{\Blue \rm Dimensions~six\Black} &\multicolumn{2}{|c|}{\Blue m_h=115\GeV\Black}&
\multicolumn{2}{|c|}{\Blue m_h=300\GeV\Black}&
\multicolumn{2}{|c}{\Blue m_h=800\GeV\Black}\\ 
\multicolumn{3}{c|}{\Blue \rm operators\Black}&
\Blue c_i=-1\Black&\Blue c_i=+1\Black&\Blue c_i=-1\Black&\Blue c_i=+1\Black&\Blue c_i=-1\Black&\Blue c_i=+1\Black\\ \hline
\hline
\Blue\Ord_{WB}&=&(H^\dagger \tau^a H) W^a_{\mu\nu} B_{\mu\nu} \Black             &9.7 & 10  &7.5 & \NO & \NO & \NO \\
\Blue\Ord_{H}&=&|H^\dagger D_\mu H|^2 \Black                                     &4.6 & 5.6 &3.4 & \NO & 2.8 & \NO \\
\Blue\Ord_{LL}&=&\frac{1}{2}(\bar{L}\gamma_\mu \tau^a L)^2 \Black                &7.9 & 6.1 &\NO & \NO & \NO & \NO \\
\Blue\Ord_{HL}' &=&i(H^\dagger D_\mu \tau^a H)(\bar{L}\gamma_\mu \tau^a L)\Black &8.4 & 8.8 &7.5 & \NO & \NO & \NO \\
\Blue\Ord_{HQ}' &=&i(H^\dagger D_\mu \tau^a H)(\bar{Q}\gamma_\mu \tau^a Q) \Black&6.6 & 6.8 &\NO & \NO & \NO & \NO \\
\Blue\Ord_{HL} &=&i (H^\dagger D_\mu H)(\bar{L}\gamma_\mu L)\Black &7.3 & 9.2& \NO & \NO& \NO & \NO \\
\Blue\Ord_{HQ} &=&i (H^\dagger D_\mu H)(\bar{Q}\gamma_\mu Q) \Black&5.8 & 3.4& \NO & \NO& \NO & \NO \\
\Blue\Ord_{HE} &=& i (H^\dagger D_\mu H)(\bar{E}\gamma_\mu E)\Black&8.2 & 7.7& \NO & \NO& \NO & \NO \\
\Blue\Ord_{HU} &=& i (H^\dagger D_\mu H)(\bar{U}\gamma_\mu U)\Black&2.4 & 3.3& \NO & \NO& \NO & \NO \\
\Blue\Ord_{HD} &=& i (H^\dagger D_\mu H)(\bar{D}\gamma_\mu D)\Black&2.1 & 2.5& \NO & \NO& \NO & \NO 

\end{array}$$
\caption{\em $95\%$ lower bounds on $\Lambda/\TeV$ for the individual operators and different values of $m_h$.
$\chi^2_{\rm min}$ is the one in the SM for $m_h> 115\GeV$.}
\end{table*}

\begin{figure*}[t]
\begin{center}
\begin{picture}(15,6.7)
\put(0,0){\includegraphics[width=15cm,height=6cm]{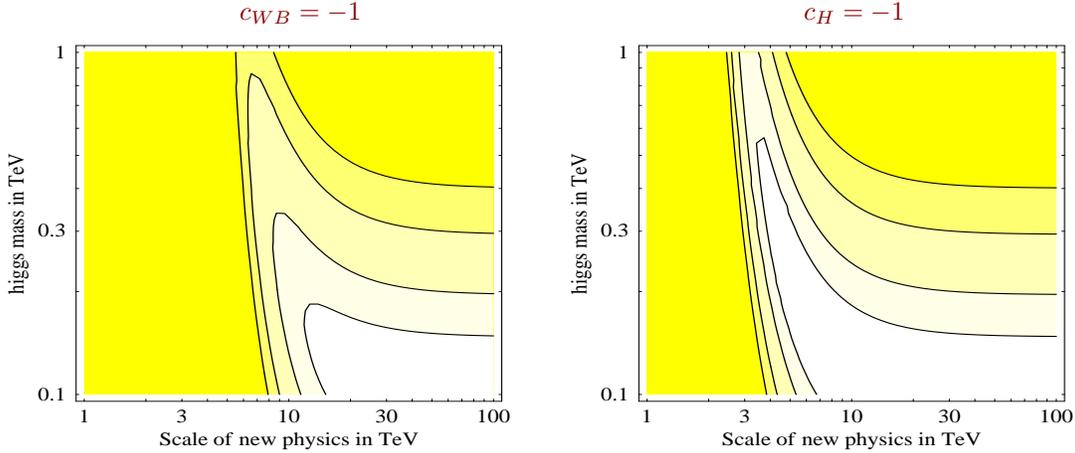}}\Red
\put(3.5,6){$c_{WB}=-1$}
\put(11,6){$c_{H}=-1$}
\Black
\end{picture}

\caption[SP]{\em Level curves of $\Delta \chi^2=\{1,2.7,6.6,10.8\}$
that correspond to $\{68\%,90\%,99\%,99.9\%\}{\rm CL}$
for the first 2 operators in table~1 ($\Ord_{WB}$ and $\Ord_{H}$)
and $c_i=-1$.
\label{fig:CP1}}
\end{center}\end{figure*}

\section{Is supersymmetry there?}
The naturalness problem of the Fermi scale, caused by the quadratic divergences in the Higgs mass, is with us since more than 20
years.
We think that a Higgs mass in the $(100\div 200)\GeV$ range and, especially, a lower bound on the scale of new physics
of about $5\TeV$ turn the naturalness problem of the Fermi scale into a clear paradox.
The loop with a top of $170\GeV$ gives a contribution to the Higgs mass
\begin{equation}\label{eq:dmh}
\delta m_h^2({\rm top}) = \frac{3}{\sqrt{2}\pi^2}G_{\rm F}m_t^2 k_{\rm max}^2 = (0.3\,k_{\rm max})^2
\end{equation}
where $k_{\rm max}$ is the maximum momentum of the virtual top.
The paradox arises if one thinks that $5\TeV$ is also a lower bound on $k_{\rm max}$, since in this case
$\delta m_h^2({\rm top})$ would exceed $(1.5~\TeV)^2$, about 100 times the indirect value of $m_h^2$.
We like to call this the ``LEP paradox'', for obvious reasons.

Supersymmetry offers a neat solution to this paradox.
A stop loop counteracts the top loop contribution to the Higgs mass, turning $k^2_{\rm max}$ of eq.\eq{dmh} into
\begin{equation}\label{eq:rep}
k^2_{\rm max}\to m_{\tilde{t}}^2\ln\frac{k_{\rm max}^2}{m_{\tilde{t}}^2}
\end{equation}
In this way a stop mass $m_{\tilde{t}}$ in the Fermi-scale range keeps the top-stop contribution to $m_h$ under control,
{\em while not undoing the success of the SM in passing the EWPT.}
This is a non trivial constraint for any possible solution of the LEP paradox.
The success of supersymmetric grand unification adds significant support to this view~\cite{GUT}.

The contrary arguments to the supersymmetric solution of the LEP paradox are of general
character.
One argument is that power divergences in field theory are not significant.
This looks problematic to us:
the top loop is there and something must be done with it.
A quadratic divergence explains the $\pi^{\pm}/\pi^0$ mass splitting 
$m_{\pi^\pm}^2-m_{\pi^0}^2\sim (\alpha_{\rm em}/4\pi) \Lambda_{\rm QCD}^2$~\cite{empimass}.
More relevant may be the observation that the cosmological constant poses another
serious unsolved problem, also related to power divergences.

Alternative physical pictures are proposed for solving the hierarchy problem
(top-colour~\cite{topcolour}, extra dimensions without supersymmetry~\cite{TeVgravity}, \ldots).
As far as we know, they all share a common problem: the lack of calculative techniques and/or of suitable conceptual developments do
not allow to address the  LEP paradox. Maybe the fundamental scale of these theories is low and the agreement of the
EWPT with the SM and a high cut-off is accidental.
Alternatively, the separation between the Higgs mass and the scale of these theories may be considerable.
In this last case, unfortunately, the related experimental signatures may become elusive.

\section{Where is supersymmetry?}
If supersymmetry solves the hierarchy problem, where is it then?
In supersymmetric models, a good approximation to the Higgs mass for moderately large $\tan\beta$ is given by
\begin{equation}\label{eq:mhQ}
m_h^2\approx \frac{3}{\sqrt{2}\pi^2}G_{\rm F}m_t^2 m_{\tilde{t}}^2 \ln\frac{Q^2}{m_{\tilde{t}}^2}.
\end{equation}
Note how this simply arises by the replacement\eq{rep} into\eq{dmh} and the identification of $k_{\rm max}$ with $Q$,
the RGE scale at which $m_h$ vanishes.
In specific models $Q$ is a function of the various parameters.

As well known, $m_h^2$ can also be computed from the quartic coupling of the Higgs potential.
Including the one loop large top corrections, one has ($\tan\beta\circa{>}4$)
\begin{equation}\label{eq:mhv}
m_h^2\approx M_Z^2 + \frac{3}{\sqrt{2}\pi^2}G_{\rm F}m_t^4\ln\frac{m_{\tilde{t}}^2}{v^2}
\end{equation}
Eq.s\eq{mhQ} and\eq{mhv} may be viewed as a relation between $Q$ and $m_{\tilde{t}}$,
graphically represented in fig.~3.

As mentioned $Q$ is a model dependent function of the various parameters, ranging from the weak scale to the Planck scale.
A random choice of the original parameters leads most often to
a point on the prolongation of the left branch of the curve in fig.~3, where $\ln(Q/ m_{\tilde{t}})\gg 1$.
However, given the correlation between stop masses and the other sparticle masses
expected in explicit models,
experiments have excluded this region, requiring that $Q\sim m_{\tilde{t}}$.

`Where is supersymmetry?' depends on the interpretation of this fact.
If it is due to an accidental fine-tuning,
it is no longer unlikely to have sparticles above a TeV due to a slightly more improbable accident.
At the same time the explanation of the LEP paradox becomes cloudy.

If instead $Q\sim m_{\tilde{t}}$ is not accidental, it is important to notice that experiments do not yet require
that we live on the right branch in fig.~3, with $Q$ very close to $m_{\tilde{t}}$.
If, for some reason, $Q\sim m_{\tilde{t}}$, so that sparticle masses are related to the weak scale by a one loop relation
(rather than by the usual tree level relation), sparticles should be around the corner.
We have recently conjectured that suitable models may exist where $Q$ is predicted to be close to the minimum in fig.~3,
where $m_{\tilde{t}}\sim 400\GeV$~\cite{NatSUSY}.
In fig.~4 we show a sampling of the
spectra expected in these models, if we also assume minimal supergravity relations between soft terms.

\begin{figure}[t]
\begin{center}
\includegraphics[height=8cm]{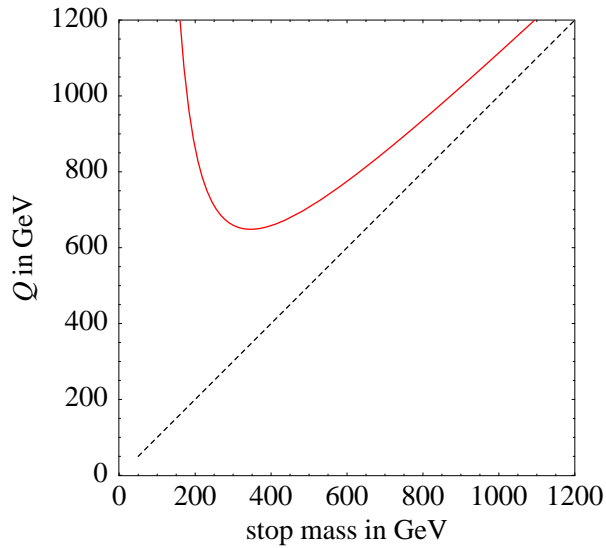}
\caption[SP]{\em Correlation between the stop mass and the RGE scale $Q$ at which $m_h(Q)$ vanishes.
\label{fig:FT}}
\end{center}\end{figure}

\begin{figure}[t]
\begin{center}
\includegraphics[height=8cm]{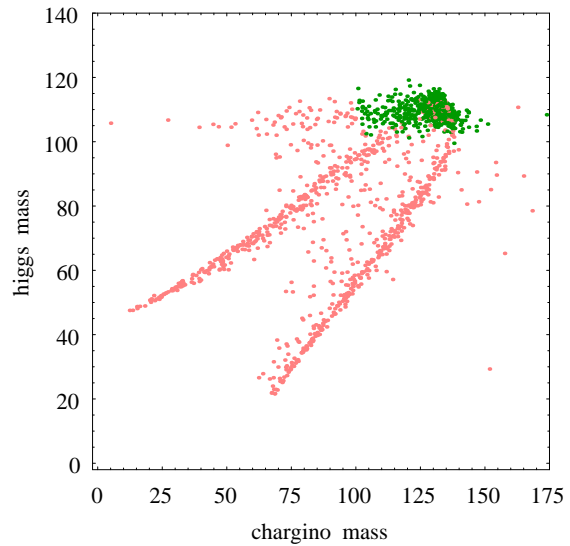}
\caption[SP]{\em Correlation between the chargino and the Higgs mass in GeV.
Sampling spectra excluded (allowed) by accelerator bounds are drawn in light red (dark green).
\label{fig:FT}}
\end{center}\end{figure}

\section{Conclusion}
A straight interpretation of the results of the EWPT, mostly performed at LEP in the last decade,
gives rise to an apparent paradox.
The EWPT indicate both a light Higgs mass $m_h\approx (100\div 200)\GeV$ and a high cut-off,
$\Lambda\circa{>}5\TeV$, with the consequence of a top loop correction to $m_h$ largely exceeding
the preferred value of $m_h$ itself.
The well known naturalness problem of the Fermi scale has gained a pure `low energy' aspect.
At present, supersymmetry at the Fermi scale is the only way we know of to attach this problem.

This way of looking at the data may be too naive.
As we said, in EWPT the SM with a light Higgs and a large cut-off can at least be faked by a fortuitous cancellation.
In any case the point is not to replace direct searches for supersymmetry or for any other kind of new physics.
Rather, we wonder if a better theoretical focus on the LEP paradox might be not without useful consequences.
Its solution, we think, is bound to give us some surprise, in a way or another.

\paragraph{Acknowledgments}
Work supported in part by the E.C. under TMR contract No. ERBFMRX--CT96--0090.
Talk given at the IVth Rencontres du Vietnam, July 2000.


\begin{thebibliography}{nn}
\bibitem{NRO-SM-FIT} \art[hep-ph/9905281]{R. Barbieri, A. Strumia}{\PL}{B462}{144}{1999}.
For the experiments, the main change is
$M_W=(80.434\pm 0.037)\GeV$.
We have updated the SM prediction for the EWPT as given to us by G. Passarino
using the program {\sc TopaZ0}.
\end{thebibliography}
\end{document}